\documentclass{INTERSPEECH2023}
\usepackage{multirow}
\usepackage{wasysym}
\usepackage{bbm}
\interspeechcameraready

\title{Ultra Dual-Path Compression For Joint Echo Cancellation And Noise Suppression}
\name{Hangting Chen, Jianwei Yu, Yi Luo, Rongzhi Gu, Weihua Li, Zhuocheng Lu, Chao Weng}
\address{
  Tencent AI Lab, Audio and Speech Signal Processing Oteam}
\email{\{erichtchen, tomasyu, oulyluo, lorrygu, weihuali, zhuochenglu, cweng\}@tencent.com}

\begin{document}

\maketitle
 
\begin{abstract}
Echo cancellation and noise reduction are essential for full-duplex communication, yet most existing neural networks have high computational costs and are inflexible in tuning model complexity. In this paper, we introduce time-frequency dual-path compression to achieve a wide range of compression ratios on computational cost. Specifically, for frequency compression, trainable filters are used to replace manually designed filters for dimension reduction. For time compression, only using frame skipped prediction causes large performance degradation, which can be alleviated by a post-processing network with full sequence modeling. We have found that under fixed compression ratios, dual-path compression combining both the time and frequency methods will give further performance improvement, covering compression ratios from 4x to 32x with little model size change. Moreover, the proposed models show competitive performance compared with fast FullSubNet and DeepFilterNet. \footnote{A demo page can be found at \url{hangtingchen.github.io/ultra\_dual\_path\_compression.github.io/}. Core source code can be found in \url{https://github.com/tencent-ailab/UltraDualPathCompression}.}

\end{abstract}

\section{Introduction}

Echo and noise are the main distractions in full-duplex voice communication systems, resulting in disruptions and difficulties in comprehending the speech of participants. With the rapid advancement of deep learning, neural networks have achieved remarkable performance on signal processing tasks. This work focuses on the joint task of acoustic echo cancellation (AEC) and noise suppression (NS) with low-complexity models to facilitate real-time communication in the real world.


Deep learning-based networks have been used to perform denoising tasks \cite{Dubey2022Icassp2D,Reddy2021Interspeech2D,Reddy2020ICASSP2D} and echo cancellation \cite{DBLP:conf/interspeech/CutlerSPLSPGBSA21,DBLP:conf/icassp/SridharCSPLGBA021,DBLP:conf/icassp/CutlerSPPGBSA22}. Despite the existence of low-complexity models for denoising, such as DeepFilterNet \cite{DBLP:conf/icassp/SchroterERM22}, tuning their computational cost is risky since the model size changes as the hyper-parameters. The Fast FullSubNet \cite{hao2022fast} explores the time and frequency compression methods but it has not balanced the compression along different axes and most compressed variants still exhibit large complexity. Compared with NS, models for the joint AEC and NS task usually take multiple signals including microphone signals, reference signals and error signals generated by linear AEC (LAEC) algorithm \cite{DBLP:conf/icassp/ZhangWSFTFX22}. Most existing models only report real time factors (RTFs) ranging from 0.05 to 0.5, lacking detailed calculation cost numbers, and still cannot meet the needs of mobile platforms \cite{DBLP:conf/icassp/ZhangWSFTFX22,DBLP:conf/icassp/WesthausenM21,DBLP:conf/icassp/ValinTHIK21}.

We choose the dual-path transformer-based full-subband network (DPT-FSNet) \cite{dangfeng} to explore compression methods for three reasons. First, the model has exhibited high wide-band perceptual evaluation of speech quality (WB-PESQ) scores on the NS task with a small number of parameters but suffers from large computational cost. Second, DPT-FSNet is conducted on complete time-frequency (T-F) feature maps, indicating its complexity being closely related to the number of frames and frequency bands. Third, the model involves a 2D convolution encoder, a dual-path transformer and a 2D convolution decoder, implying that compression methods should be applicable to different modules.




To address the issues of real-time applications, time and frequency compression methods are explored for model complexity reduction. We propose Mel scale-based trainable compression and frame-skipped prediction followed by a post-processing network for each axis, respectively. The Mel scale-based trainable filters are used for frequency compression instead of using manually fixed filters on the amplitude spectrum; the frame-skipped prediction is able to compress the time sequence length but suffers from unmatched masks. An additional light post-processing network with low complexity can alleviate the degradation by utilizing full-sequence modeling. Moreover, under fixed compression ratios,  appropriate integration of time and frequency methods, named dual-path compression, can obtain minimized performance degradation by alleviating the large information loss caused by excessive compression.

Experimental results show that Mel scale-based trainable compression outperforms classic equivalent rectangular bandwidth (ERB) and Mel filters. In time compression, post-processing networks obtain more than 0.2 WB-PESQ improvement from 4x to 32x compression ratio. Furthermore, the dual-path compression achieves approximately 0.1 WB-PESQ higher than single-path compression on 8x and 16x ratios.

We claim that the proposed models can cover a wide range of computational complexity without reducing model sizes and exhibit competitive performance under the joint tasks of AEC and NS. Specifically, the multiply-accumulate operations per second (MACs/s) of the presented models cover the range from 100M to 2000M.\footnote{A model with 160M MACs/s has an RTF of 0.01 using Intel (R) Core (TM) i7-7700 @ 3.60GHz in the real-world application.} Moreover, we compared the proposed models with DeepFilterNet \cite{DBLP:conf/icassp/SchroterERM22} and Fast FullSubNet \cite{hao2022fast}. The results show that our models outperform fast FullSubNet and achieve similar performance as DeepFilterNet but with more flexible compression ratios and fewer parameters.



\section{Ultra low-complexity DPT-FSNet}

\subsection{Online DPT-FSNet}\label{sec:model}

\begin{figure}
\centerline{\resizebox{0.37\textwidth}{!}{\includegraphics[width=\columnwidth]{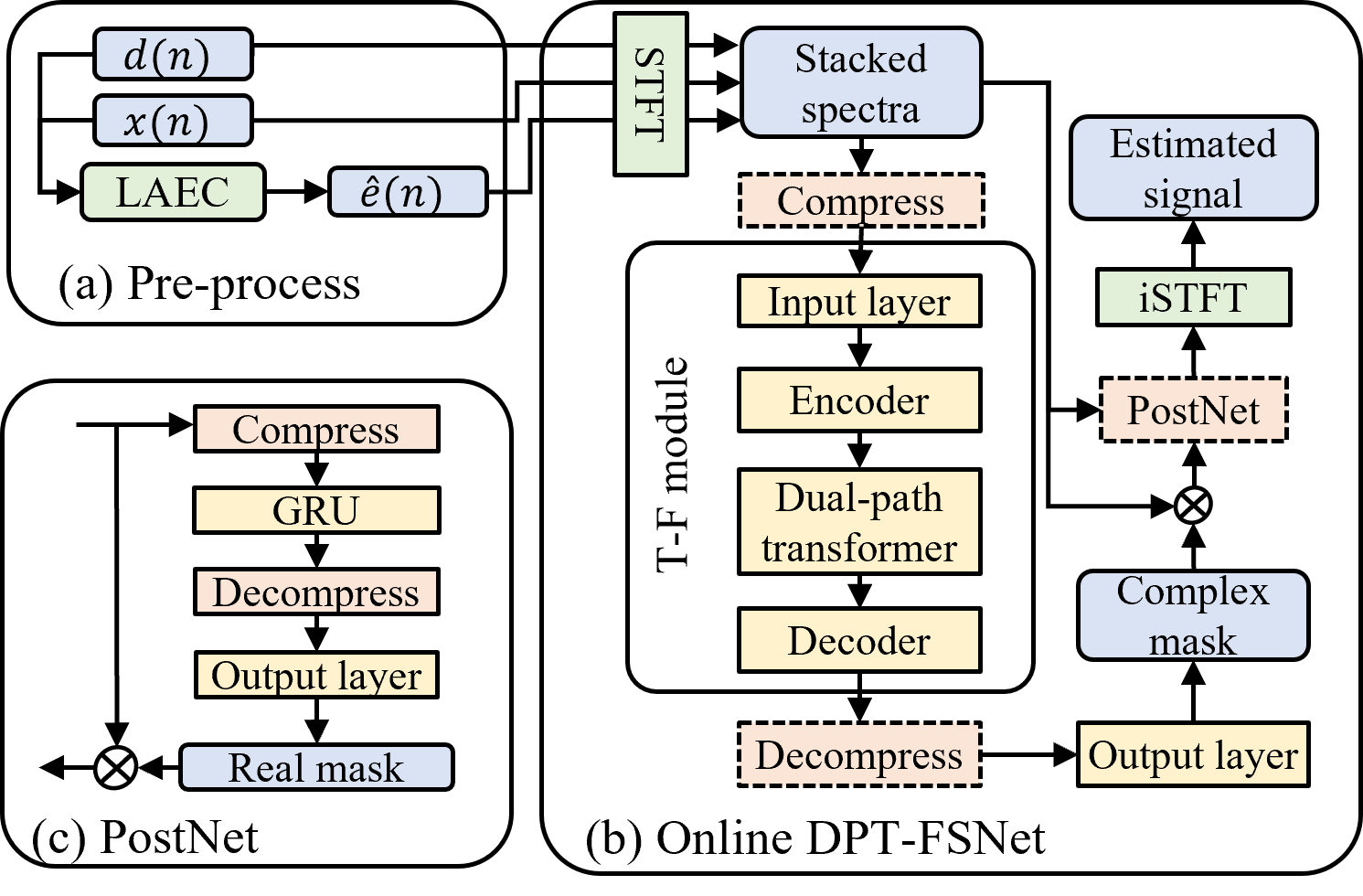}}}
\caption{A pipeline for joint echo and noise suppression. The boxes with dashed lines are optional modules for compression.}
\label{fig1}
\end{figure}

The offline DPT-FSNet processes the T-F feature with a 2D convolutional encoder/decoder and a dual-path transformer (Fig.\ref{fig1}) \cite{dangfeng}. To enable streaming processing with reduced computational cost, the online DPT-FSNet was redesigned. First, the input layer accepts multiple input signals, including the microphone signal $d$, the reference signal $x$ and the error signal $\hat{e}$ generated by LAEC \cite{DBLP:conf/icassp/KuechME14}. The output layer generates individual masks for each input, and the final output is the summation of all filtered signals. Second, to reduce computational complexity, long short-term memory layers in the dual-path transformer were replaced with a single gated recurrent unit (GRU) placed after the first attention layer. Linear attention is adopted as its linear complexity and memory use \cite{DBLP:conf/icml/KatharopoulosV020}. Third, the subband attention and GRU layers are unidirectional for online processing, while the fullband ones are bidirectional.





The input spectra are fed into an input layer, a T-F module and an output layer sequentially. The compression and decompression modules are inserted before the input and output layers, respectively. The input layer together with the compression transforms stacked spectra with dimension $2C\times T\times F$ into $E\times T\times F$, where $T$ and $F$ are numbers of frames and frequencies, $E$ is the feature dimension for each T-F bin and $2C$ is the stacked real and imaginary channels. The output layer together with the decompression accepts a feature with dimension $E\times T\times F$ and generates the complex masks with dimension $2C\times T\times F$. Both the input and output layers adopt $1\times 1$ convolution. We present several methods of time and frequency compression that aim to reduce the computational cost with minimized performance degradation and small model sizes (e.g., parameters less than 500K are feasible for current devices).


\begin{figure}
\centerline{\resizebox{0.35\textwidth}{!}{\includegraphics[width=\columnwidth]{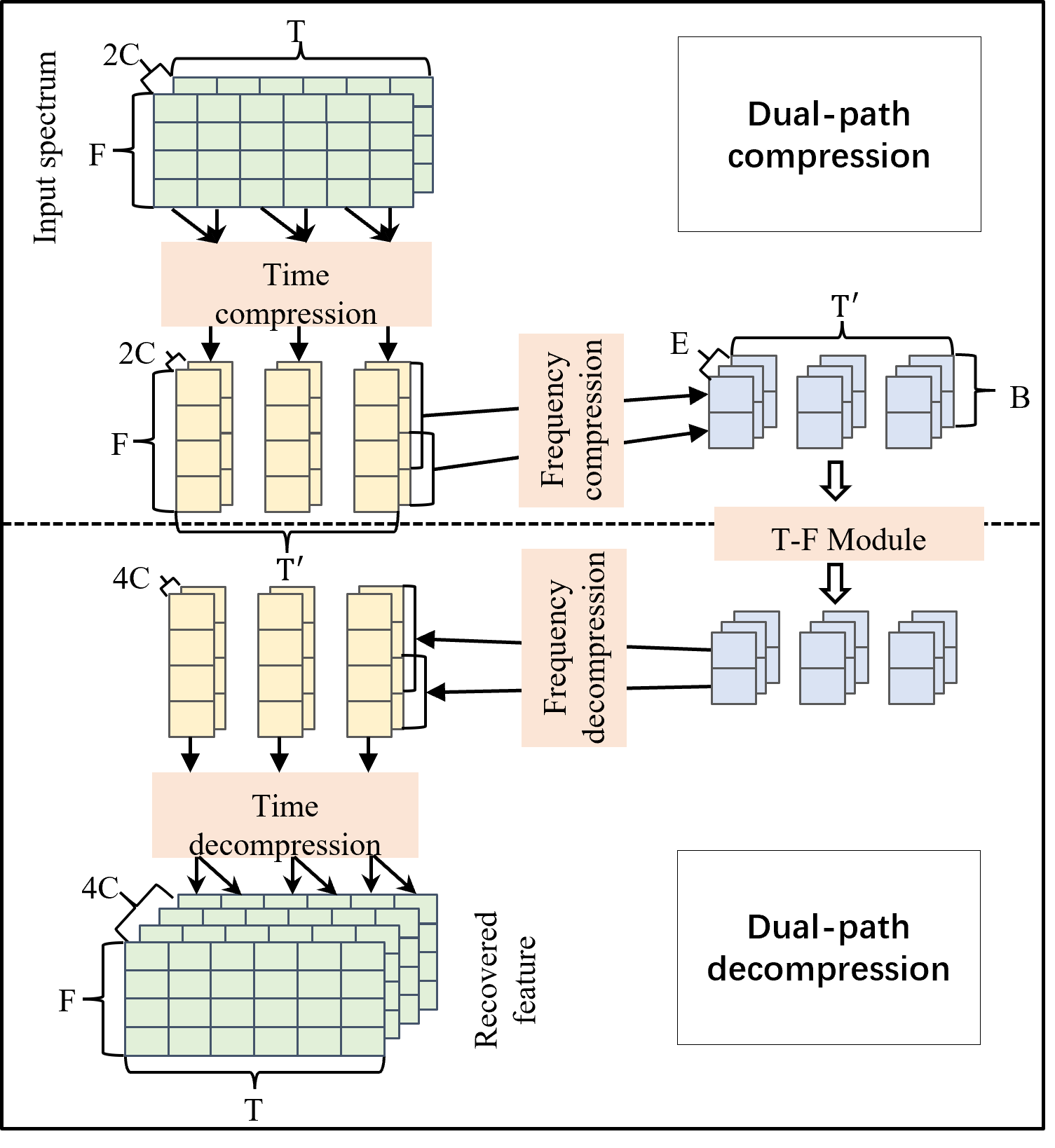}}}
\caption{Dual-path compression and decompression.}
\label{fig2}
\end{figure}

\subsection{Frequency compression}

\subsubsection{Manually fixed filters}

We explore fixed triangle filters based on ERB \cite{moore1983suggested} and Mel scale \cite{stevens1937scale}. The compressed features are calculated by,
\begin{equation}
Z[c,t,b]={\rm log}(\sum_{f={\rm low}[b]}^{{\rm high}[b]}|X[c,t,f]|\times W_c[b,f]),
\end{equation}
where $|X[c,t,f]|$ is the modulus of the complex spectrum with signal index $c$, frame index $t$ and frequency index $f$, $W[b,f]$ is the weight for $b$th filter on frequency $f$, ${\rm low}[b]$ and ${\rm high}[b]$ are the start and end frequencies of $b$th filter, $Z$ is the output feature with dimension $C\times T\times B$. The input layer with $1\times1$ convolution transforms the compressed feature $Z$ into $E\times T\times B$ for the following blocks. For decompression, we use Moore–Penrose inverse of filter weight $W$ to linearly transform the feature into dimension $E\times T\times F$ for the output layer. 


\subsubsection{Trainable filters}

Different from manually fixed filters, complex spectrum-based compression uses trainable transformations. We follow specific scales to split bands. The real and imaginary parts of complex spectra are stacked and transformed into the high-level feature space using a linear transformation as follows:
\begin{equation}
    Z[:,t,b]={\rm Flatten}(X[0:C,t,{\rm low}[b]:{\rm high}[b]])\times W,
\end{equation}
where $W \in \mathbbm{R}^{(\triangle B[b]\times C)\times E}$ converts spectra into $E\times T\times B$, $\triangle B[b]={\rm high}[b]-{\rm low}[b]$. With trainable compression, the $1\times 1$ convolution in the input layer is redundant and should be removed. For decompression, another linear transformation is used to decompress the $E$-dimensional vector into dimension $4C\times \triangle B[b]$ for the output layer processing, where we use $4C\times \triangle B[b]$ instead of $E\times \triangle B[b]$ to save the model size. 

\subsection{Time compression}

We use skip prediction and a post-processing network for time compression. The skip prediction module concatenates previous frames and predicts the current mask. Meanwhile, a post-processing network is employed to reduce the performance degradation caused by unmatched output. Compared with Fast FullSubNet \cite{hao2022fast} compressing the time axis for subband model only, our compressed feature is used for both fullband and subband parts. Moreover, our post-processing network had only 67K parameters and 15M MACs/s.

\subsubsection{Skip prediction}

Fig.\ref{fig2} illustrates the processing procedure for skip prediction. At each $r$ frame interval, the compression module accepts the stacked current and history frames and uses a linear transformation to generate features with dimension $E\times T'\times F$, where $T'=T/r$. For decompression, we use the generated features for the current frame and copy it $r-1$ times for future frames.

\subsubsection{Post-processing network}

Skip prediction results in significant performance degradation due to unmatched output. To alleviate this issue, we employ a light post-processing module for full-sequence processing. Fig.\ref{fig1}(c) shows the architecture of the post-processing network, which consists of a feature compression module, a 1-layer Gated Recurrent Unit (GRU), a feature decompression module, and an output layer. With the limited modeling ability of the 1-layer GRU, we use the log power spectra of both the error signal $\hat{e}$ and the signal from the previous stage as inputs. The input feature is of dimension $2\times T\times F$. We conduct the frequency compression with ratio 2 to save computational cost. The output layer employs stacked $1\times 1$ convolutions, linear transforms, and a sigmoid activation to predict real-valued masks.


\subsection{Dual-path compression}

Dual-path compression combines time and frequency ways to achieve ultra-computational cost reduction. The search space increases linearly corresponding to the exponentially increasing compression ratio. For example, the search space is 3 regarding to compression ratio 4 with base ratio 2 (i.e. T-F compression ratios of $1\times 4$, $2\times 2$ and $4\times 1$). The best compression combination is unknown, so grid search is used to obtain the optimal combination. The orders of time and frequency compression exhibit little performance difference. Thus we adopt time compression and then frequency compression followed by frequency decompression and time decompression (Fig.\ref{fig2}).

\subsection{Discussion and analysis}

Reducing model complexity can be easily achieved by shrinking model size. For example, a smaller $E$ will bring a lower computational complexity as well as the model size. Since the online DPT-FSNet already has a small size ($\sim$100K), shrinking $E$ will cause significant performance degradation (Fig.\ref{fig3}). 

The proposed compression can cover a wide range of compression ratios straightforwardly, which can be deployed on most neural network-based front-end models. Nonetheless, the compression and decompression modules have a certain amount of parameters and computational complexity, resulting in lowering the complexity hard in large compression ratios. We will leave compression ratios larger than 32 for future work.

\section{Experimental setup}\label{sec:exp}

\subsection{Dataset}

We used simulated audios to validate the proposed methods. Clean audios from Librispeech \cite{DBLP:conf/icassp/PanayotovCPK15} \textit{train-clean-100} and \textit{train-clean-360} were used to generate training and validation set while \textit{test-clean} was used to generate evaluation set. The noise audios from DNS-Challenge \cite{DBLP:conf/interspeech/ReddyDKNGCBGA021} were split for training, validation and evaluation set, individually. The echos in the training and validation sets were simulated using clean audios and room impulse responses (RIRs) taken from a published dataset \cite{DBLP:conf/icassp/KoPPSK17} while echos in the evaluation set were real-recorded taken from AEC-challenge \cite{DBLP:conf/icassp/CutlerSPPGBSA22} which covered various devices and signal delays \cite{DBLP:conf/icassp/ZhangWSFTFX22}. The simulated training, validation and evaluation set had durations of 530h, 10h and 10h. All audios were sampled in 16 kHz with lengths ranging from 9 to 10 seconds.

In detail, we considered 3 scenarios including far-end single talk (ST-FE), near-end single talk (ST-NE) and double talk (DT). The training and validation set shared the same data simulation configuration. Both the signal-to-echo ratio (SER) and the signal-to-noise ratio (SNR) were uniformly sampled from $-5$ to 15 dB. The near-end speakers were not present in 10\% of the data while the far-end speakers were not present in 25\% of the data. Approximately 90\% audios were noisy while 10\% only contains speech. The evaluation set had SERs and SNRs of $-5$ dB, 5 dB, 15 dB and $+\infty$. SER equal to $+\infty$ corresponded to ST-NE scenario while SER equal to $-\infty$ corresponded to ST-FE scenario. These settings ensured that the performance improvement could be verified under different environments. Due to the limited pages, we report the average performance in the aforementioned 3 scenarios.

\subsection{Performance metrics}

The ST-FE scenario was evaluated by echo return loss enhancement (ERLE). ST-NE and DT scenarios were evaluated by scale-invariant signal-to-noise ratio (SI-SNR)  \cite{Luo2019ConvTasNetSI}, WB-PESQ \cite{DBLP:conf/icassp/RixBHH01} and short-time objective intelligibility (STOI) \cite{DBLP:conf/icassp/TaalHHJ10}. The widely used C version of WB-PESQ is adopted\footnote{https://www.itu.int/rec/T-REC-P.862/en}.

\subsection{Model setup}

\begin{figure}
\resizebox{0.4\textwidth}{!}{
\centerline{\includegraphics[width=\columnwidth]{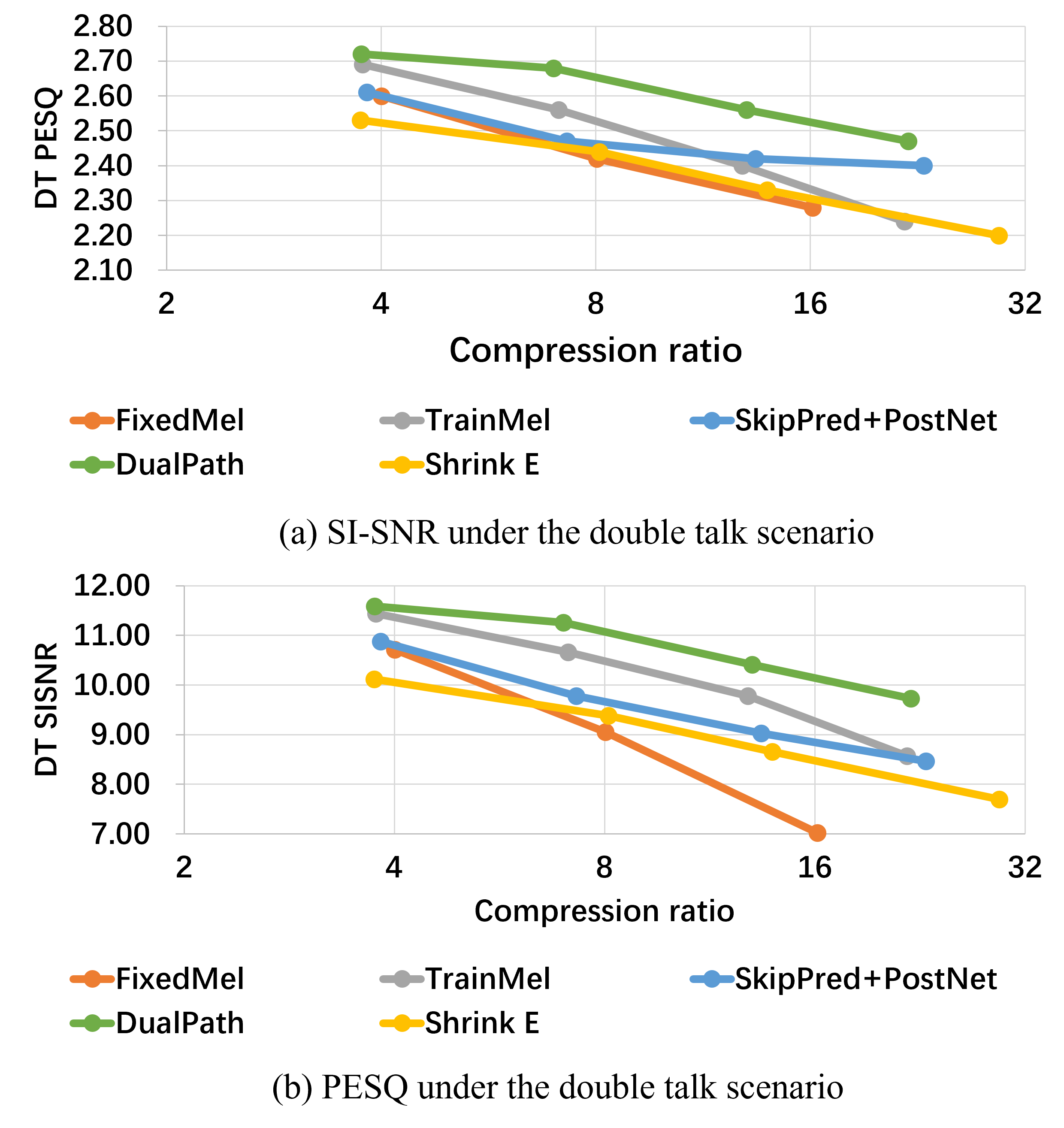}}}
\caption{Performance of different compression methods under the double talk scenario. }
\label{fig3}
\end{figure}

The STFT and iSTFT used a window size of 20 ms and a stride size of 10 ms. A state-space based linear filter was first used to estimate error signal $\hat{e}$ \cite{DBLP:journals/sigpro/EnznerV06,DBLP:conf/icassp/KuechME14}. The number of convolution layers was set to 2 in the encoder and decoder. The dual-path GRU number was set to 1 while the dual-path attention layer number was set to 4. The feature dimension $E$ was set to 48. The whole model had a parameter size of 109K and a computational cost of 1822M MACs/s. The hyper-parameter $B$ in PostNet was fixed to $80$. The change of $B$ had a limited effect on the parameter size and computational cost because compression, decompression and output layer occupied a non-negligible ratio.

\begin{table*}[!t]
\centering
\caption{Performance of time compression, frequency compression and dual-path compression. \label{tab:table2}}
\resizebox{0.85\textwidth}{!}{
\begin{tabular}{l|c|l|c|c|ccc|ccc|c}
\toprule\hline
\multirow{2}{*}{Method} & \multirow{2}{*}{T$\times$F Ratio}  & \multirow{2}{*}{\# Param.} & \multirow{2}{*}{MACs/s} & Compression & \multicolumn{3}{c|}{DT}                                & \multicolumn{3}{c|}{ST-NE}                             & ST-FE          \\
                  &     & &    &       ratio         & \multicolumn{1}{c}{SI-SNR} & \multicolumn{1}{c}{WB-PESQ} & \multicolumn{1}{c|}{STOI} & \multicolumn{1}{c}{SI-SNR} & \multicolumn{1}{c}{WB-PESQ} & \multicolumn{1}{c|}{STOI} & ERLE           \\ \hline
Fast Fullsubnet       &  -           & 7601K  & 1433M  & -     & 11.33 & 2.68 & 87.05 & 13.27 & 2.84 & 89.31 & 40.66 \\
DeepFilterNet         & -            & 2000K  & 289M   & -     & 10.96 & 2.68 & 86.90 & 12.97 & 2.83 & 88.95 & 44.49 \\
Uncompressed          & $1\times 1$  & 109K   & 1822M  & 1.0   &  \textbf{12.14}   & \textbf{2.78} & \textbf{87.96}  & \textbf{13.83}  & \textbf{2.91} & \textbf{89.70}  & \textbf{46.82}          \\ \hline
FixedERB  & $1\times 2$  & 109K   & 910M   & 2.0 & 10.10  & 2.54 & 85.14 & 11.83   & 2.70 & 87.39  & 43.99          \\
FixedMel  & $1\times 2$  & 109K   & 910M   & 2.0 & 11.18   &  2.69  & 86.82 & 12.88 &  2.83 & 88.71 &  44.33          \\
TrainMel  & $1\times 2$  & 413K   & 937M  &  1.9  & \textbf{11.89}    & 2.73 & \textbf{87.73} & \textbf{13.55}   & 2.87   & \textbf{89.55} & \textbf{44.68}         \\
SkipPred  & $2\times 1$  & 109K   & 917M   & 2.0   & 11.31  & 2.68   & 86.80 & 13.05   & 2.81  & 88.67  & 40.44          \\ 
\ +PostNet       & $2\times 1$  & 177K   & 931M   &  2.0  & 11.86  & \textbf{2.75}  & 87.66 & 13.53   &  \textbf{2.89}  & 89.40 & 43.83               \\  \hline
FixedERB  & $1\times 4$  & 109K   & 455M & 4.0  &  8.88   & 2.40  & 83.17 & 10.56   & 2.56  &  85.71 & 41.64          \\
FixedMel  & $1\times 4$  & 109K   & 455M & 4.0  &  10.71  & 2.60  & 85.95 & 12.27   & 2.72   & 87.81 & 41.97          \\
TrainMel  & $1\times 4$  & 408K   & 484M & 3.8   &  11.44  & 2.69  & 86.97 & 13.23   & 2.83  &  89.01 & 42.13          \\
SkipPred  & $4\times 1$  & 110K & 462M &  3.9  & 9.83   & 2.39  & 83.46 & 11.48   &  2.54  & 85.86 &  40.26         \\
\ +PostNet       & $4\times 1$  & 177K & 477M &  3.8  &  10.88  &  2.61  & 85.98 &  12.61  & 2.76  & 88.05  &  38.35         \\ 
DualPath       & $2\times 2$  & 481K & 486M & 3.7 &  \textbf{11.59}  & \textbf{2.72}   & \textbf{87.44} &  \textbf{13.49} &  \textbf{2.88}  & \textbf{89.26} & \textbf{42.63}        \\  \hline
FixedERB  & $1\times 8$  & 109K   & 227M & 8.0  &  7.96   & 2.30  & 81.71  & 9.67    & 2.46  &  84.55 & 38.04          \\
FixedMel  & $1\times 8$  & 109K   & 227M & 8.0   &  9.05  & 2.42  & 83.65 & 10.66   & 2.56  &  86.04 & 41.52          \\
TrainMel  & $1\times 8$  & 398K   & 257M & 7.1  &  10.66  & 2.56  & 85.36 & 12.56   & 2.73  &  87.84 & 41.10          \\
SkipPred  & $8\times 1$  & 111K & 245M & 7.4   & 7.64   & 2.14  & 78.26  & 9.44   & 2.32  & 81.74  & 39.42          \\
\ +PostNet       & $8\times 1$  & 178K & 250M & 7.3   &  9.78  & 2.47  & 84.35 & 11.86   & 2.64  &  87.15 & 36.64          \\ 
DualPath       & $2\times 4$  & 476K & 261M & 7.0   & \textbf{11.26}   & \textbf{2.68}  &  \textbf{86.82} &  \textbf{13.05}  & \textbf{2.82}  &  \textbf{88.83} &  \textbf{42.06}         \\ \hline
FixedERB  & $1\times 16$ & 109K   & 113M & 16.1  &  6.79  & 2.14  & 79.25 & 8.47   & 2.31  & 82.72  & 36.50         \\
FixedMel  & $1\times 16$ & 109K   & 113M & 16.1  &  7.02  & 2.28  & 81.45 & 8.25   & 2.44  & 84.17  & 38.11          \\
TrainMel  & $1\times 16$ & 381K   & 142M & 12.8  &  9.78  & 2.40  & 83.37 & 11.76   & 2.58  &  86.49 & \textbf{40.48}          \\
SkipPred  & $16\times 1$  & 113K & 121M &  15.1  &  5.72  & 2.03  & 74.73 & 7.77   &  2.23  & 79.35 & 34.70          \\
\ +PostNet       & $16\times 1$  & 181K & 136M &  13.4  &  9.03  & 2.42  & 83.68 & 11.28   & 2.60  & 86.95  &  35.73         \\ 
DualPath       & $4\times 4$  & 477K & 140M & 13.0   & \textbf{10.41}   & \textbf{2.56 } & \textbf{85.73} &  \textbf{12.44}  &  \textbf{2.73 } & \textbf{88.07} &  39.34         \\   \hline
FixedERB  & $1\times 32$ & 109K   &  57M & 32.0  &  5.31  & 1.96  & 76.26 & 6.95   & 2.13  & 80.13  & 31.16          \\
FixedMel  & $1\times 32$ & 109K   &  57M & 32.0  &  4.27  & 2.19  & 78.76 & 4.98   & 2.35   & 81.95 & 34.35          \\
TrainMel  & $1\times 32$ & 354K   & 84M & 21.7  &  8.57  & 2.24  & 81.16 & 10.61   & 2.44  & 84.74  & 34.46          \\
SkipPred  & $32\times 1$  & 118K & 65M & 28.0   &  4.51  & 2.02  & 74.95 &  6.72  &  2.22  & 79.84 & 30.65          \\
\ +PostNet       & $32\times 1$  & 185K & 79M & 23.1   & 8.47   & 2.40  & 83.19 & 10.89   & 2.58  & 86.69  &  32.81         \\ 
DualPath       & $4\times 8$  & 467K & 83M &  22.0  &  \textbf{9.73}  & \textbf{2.47}  & \textbf{84.72} & \textbf{12.04}   & \textbf{2.66}  & \textbf{87.49}  &  \textbf{38.54}         \\   \hline\bottomrule
\end{tabular}}
\end{table*}

Models were trained using Adam optimizer on an 8-GPU machine with a batch size of $80$. Each model was optimized with 105K iterations. The results were averaged using saved checkpoints trained in 95K, 100K and 105K iterations to reduce the performance fluctuation.

\section{Results and discussion}\label{sec:results}


Different compression methods under the same DPT-FSNet architecture were compared, as listed in Table \ref{tab:table2}. Frequency compression was evaluated using manually fixed filters with ERB scale (FixedERB), with Mel scale (FixedMel), and trainable filters with Mel scale (TrainMel). FixedMel and FixedERB were widely used for model complexity reduction \cite{hao2022fast,DBLP:conf/icassp/SchroterERM22}. The results showed that WB-PESQ and ERLE scores of FixedMel outperformed those of FixedERB across all compression ratios. However, the SI-SNR of FixedMel was lower than FixedERB on large compression ratios (e.g., 16x compression). The reason might be that the SI-SNR calculation treated all frequencies equally while the Mel scale emphasized the low-frequency part more than the high-frequency part. TrainMel compression raised a WB-PESQ improvement of more than $0.1$ on 8x and 16x ratios. TrainMel increases in approximately 300K parameters and 30M MACs/s. Nevertheless, the model parameters remained under 500K. Meanwhile, compression using trainable filters could achieve similar performance with much lower computational cost, for example, the performance of 4x TrainMel is close to that of 2x FixedMel.

On time compression, we compared skip prediction (SkipPred) and the one followed by the post-processing network ($+$PostNet). SkipPred exhibited a significant performance degradation compared to TrainMel. However, PostNet largely alleviated this performance drop. At high compression ratios, WB-PESQ scores of time compression outperformed those of frequency compression as short frequency sequences lost much information when the frequency compression ratio was large.

On dual-path compression, we listed the combination of time and frequency compression to achieve optimal metrics. Compared with trainable compression with Mel scale and time compression with the post-processing network, the dual-path compression obtained higher WB-PESQ and SI-SNR scores. For example, under 16x compression, TrainMel, SkipPred+PostNet and DualPath had similar MACs while DualPath outperformed the others by more than 0.5 dB in SI-SNR and 0.1 in WB-PESQ. One drawback of dual-path compression was that its ERLE did not outperform those of TrainMel. This was attributed to the post-processing network in time compression, which caused low ERLEs and was brought to dual-path compression. We will leave this problem in future work.

Compression modules occupied a part of the computational cost, making the fair comparison tough. We exhibited several representative compression methods with computational cost versus SI-SNR and WB-PESQ under the DT scenario (Fig.\ref{fig3}). We have found that dual-path compression outperformed other methods consistently. Moreover, we compare the models with fast FullSubNet and DeepFilterNet with stacked spectra as the input. We set the down-sampling factor as 8 in fast FullSubNet. TrainMel with compression ratio 2 outperformed fast FullSubNet while having lower MACs/s and smaller model sizes. We set the convolution channel number to 48 in DeepFilterNet to match the model complexity with DualPath($2\times4$). The DeepFilterNet and DualPath($2\times4$) achieved similar performance, yet our model occupied only 1/4 storage of the DeepFilterNet.

\section{Conclusion}\label{sec:conclusion}

In this paper, we investigate dual-path compression for joint echo and noise suppression. We propose an online DPT-FSNet and explore various compression methods along both the frequency and time axes. Trainable compression and frame skip prediction with a post-processing network yielded superior performance compared to other methods along frequency and time axes, respectively. Dual-path compression further improved the performance by appropriately combining time and frequency compression. In future work, we intend to address the ERLE degradation caused by time compression and explore more advanced ways to minimize performance degradation.

\bibliographystyle{IEEEtran}
\bibliography{mybib}

\begin{thebibliography}{10}
\providecommand{\url}[1]{#1}
\csname url@samestyle\endcsname
\providecommand{\newblock}{\relax}
\providecommand{\bibinfo}[2]{#2}
\providecommand{\BIBentrySTDinterwordspacing}{\spaceskip=0pt\relax}
\providecommand{\BIBentryALTinterwordstretchfactor}{4}
\providecommand{\BIBentryALTinterwordspacing}{\spaceskip=\fontdimen2\font plus
\BIBentryALTinterwordstretchfactor\fontdimen3\font minus
  \fontdimen4\font\relax}
\providecommand{\BIBforeignlanguage}[2]{{%
\expandafter\ifx\csname l@#1\endcsname\relax
\typeout{** WARNING: IEEEtran.bst: No hyphenation pattern has been}%
\typeout{** loaded for the language `#1'. Using the pattern for}%
\typeout{** the default language instead.}%
\else
\language=\csname l@#1\endcsname
\fi
#2}}
\providecommand{\BIBdecl}{\relax}
\BIBdecl

\bibitem{Dubey2022Icassp2D}
H.~Dubey, V.~Gopal, R.~Cutler, A.~Aazami, S.~Matusevych, S.~Braun, S.~E.
  Eskimez, M.~Thakker, T.~Yoshioka, H.~Gamper, and R.~Aichner, ``Icassp 2022
  deep noise suppression challenge,'' \emph{ICASSP 2022 - 2022 IEEE
  International Conference on Acoustics, Speech and Signal Processing
  (ICASSP)}, pp. 9271--9275, 2022.

\bibitem{Reddy2021Interspeech2D}
C.~K.~A. Reddy, H.~Dubey, K.~Koishida, A.~A. Nair, V.~Gopal, R.~Cutler,
  S.~Braun, H.~Gamper, R.~Aichner, and S.~Srinivasan, ``Interspeech 2021 deep
  noise suppression challenge,'' in \emph{Interspeech}, 2021.

\bibitem{Reddy2020ICASSP2D}
C.~K.~A. Reddy, H.~Dubey, V.~Gopal, R.~Cutler, S.~Braun, H.~Gamper, R.~Aichner,
  and S.~Srinivasan, ``Icassp 2021 deep noise suppression challenge,''
  \emph{ICASSP 2021 - 2021 IEEE International Conference on Acoustics, Speech
  and Signal Processing (ICASSP)}, pp. 6623--6627, 2020.

\bibitem{DBLP:conf/interspeech/CutlerSPLSPGBSA21}
\BIBentryALTinterwordspacing
R.~Cutler, A.~Saabas, T.~P{\"{a}}rnamaa, M.~Loide, S.~Sootla, M.~Purin,
  H.~Gamper, S.~Braun, K.~S{\o}rensen, R.~Aichner, and S.~Srinivasan,
  ``{INTERSPEECH} 2021 acoustic echo cancellation challenge,'' in
  \emph{Interspeech 2021, 22nd Annual Conference of the International Speech
  Communication Association, Brno, Czechia, 30 August - 3 September 2021},
  H.~Hermansky, H.~Cernock{\'{y}}, L.~Burget, L.~Lamel, O.~Scharenborg, and
  P.~Motl{\'{\i}}cek, Eds.\hskip 1em plus 0.5em minus 0.4em\relax {ISCA}, 2021,
  pp. 4748--4752. [Online]. Available:
  \url{https://doi.org/10.21437/Interspeech.2021-1870}
\BIBentrySTDinterwordspacing

\bibitem{DBLP:conf/icassp/SridharCSPLGBA021}
\BIBentryALTinterwordspacing
K.~Sridhar, R.~Cutler, A.~Saabas, T.~P{\"{a}}rnamaa, M.~Loide, H.~Gamper,
  S.~Braun, R.~Aichner, and S.~Srinivasan, ``{ICASSP} 2021 acoustic echo
  cancellation challenge: Datasets, testing framework, and results,'' in
  \emph{{IEEE} International Conference on Acoustics, Speech and Signal
  Processing, {ICASSP} 2021, Toronto, ON, Canada, June 6-11, 2021}.\hskip 1em
  plus 0.5em minus 0.4em\relax {IEEE}, 2021, pp. 151--155. [Online]. Available:
  \url{https://doi.org/10.1109/ICASSP39728.2021.9413457}
\BIBentrySTDinterwordspacing

\bibitem{DBLP:conf/icassp/CutlerSPPGBSA22}
\BIBentryALTinterwordspacing
R.~Cutler, A.~Saabas, T.~P{\"{a}}rnamaa, M.~Purin, H.~Gamper, S.~Braun,
  K.~S{\o}rensen, and R.~Aichner, ``{ICASSP} 2022 acoustic echo cancellation
  challenge,'' in \emph{{IEEE} International Conference on Acoustics, Speech
  and Signal Processing, {ICASSP} 2022, Virtual and Singapore, 23-27 May
  2022}.\hskip 1em plus 0.5em minus 0.4em\relax {IEEE}, 2022, pp. 9107--9111.
  [Online]. Available: \url{https://doi.org/10.1109/ICASSP43922.2022.9747215}
\BIBentrySTDinterwordspacing

\bibitem{DBLP:conf/icassp/SchroterERM22}
\BIBentryALTinterwordspacing
H.~Schr{\"{o}}ter, A.~N. Escalante{-}B., T.~Rosenkranz, and A.~Maier,
  ``Deepfilternet: {A} low complexity speech enhancement framework for
  full-band audio based on deep filtering,'' in \emph{{IEEE} International
  Conference on Acoustics, Speech and Signal Processing, {ICASSP} 2022, Virtual
  and Singapore, 23-27 May 2022}.\hskip 1em plus 0.5em minus 0.4em\relax
  {IEEE}, 2022, pp. 7407--7411. [Online]. Available:
  \url{https://doi.org/10.1109/ICASSP43922.2022.9747055}
\BIBentrySTDinterwordspacing

\bibitem{hao2022fast}
X.~Hao and X.~Li, ``Fast fullsubnet: Accelerate full-band and sub-band fusion
  model for single-channel speech enhancement,'' \emph{arXiv preprint
  arXiv:2212.09019}, 2022.

\bibitem{DBLP:conf/icassp/ZhangWSFTFX22}
\BIBentryALTinterwordspacing
S.~Zhang, Z.~Wang, J.~Sun, Y.~Fu, B.~Tian, Q.~Fu, and L.~Xie, ``Multi-task deep
  residual echo suppression with echo-aware loss,'' in \emph{{IEEE}
  International Conference on Acoustics, Speech and Signal Processing, {ICASSP}
  2022, Virtual and Singapore, 23-27 May 2022}.\hskip 1em plus 0.5em minus
  0.4em\relax {IEEE}, 2022, pp. 9127--9131. [Online]. Available:
  \url{https://doi.org/10.1109/ICASSP43922.2022.9746733}
\BIBentrySTDinterwordspacing

\bibitem{DBLP:conf/icassp/WesthausenM21}
\BIBentryALTinterwordspacing
N.~L. Westhausen and B.~T. Meyer, ``Acoustic echo cancellation with the
  dual-signal transformation {LSTM} network,'' in \emph{{IEEE} International
  Conference on Acoustics, Speech and Signal Processing, {ICASSP} 2021,
  Toronto, ON, Canada, June 6-11, 2021}.\hskip 1em plus 0.5em minus 0.4em\relax
  {IEEE}, 2021, pp. 7138--7142. [Online]. Available:
  \url{https://doi.org/10.1109/ICASSP39728.2021.9413510}
\BIBentrySTDinterwordspacing

\bibitem{DBLP:conf/icassp/ValinTHIK21}
\BIBentryALTinterwordspacing
J.~Valin, S.~V. Tenneti, K.~Helwani, U.~Isik, and A.~Krishnaswamy,
  ``Low-complexity, real-time joint neural echo control and speech enhancement
  based on percepnet,'' in \emph{{IEEE} International Conference on Acoustics,
  Speech and Signal Processing, {ICASSP} 2021, Toronto, ON, Canada, June 6-11,
  2021}.\hskip 1em plus 0.5em minus 0.4em\relax {IEEE}, 2021, pp. 7133--7137.
  [Online]. Available: \url{https://doi.org/10.1109/ICASSP39728.2021.9414140}
\BIBentrySTDinterwordspacing

\bibitem{dangfeng}
F.~Dang, H.~Chen, and P.~Zhang, ``Dpt-fsnet: Dual-path transformer based
  full-band and sub-band fusion network for speech enhancement,'' in \emph{2022
  {IEEE} International Conference on Acoustics, Speech and Signal Processing,
  {ICASSP} 2022}.\hskip 1em plus 0.5em minus 0.4em\relax {IEEE}, 2022.

\bibitem{DBLP:conf/icassp/KuechME14}
\BIBentryALTinterwordspacing
F.~Kuech, E.~Mabande, and G.~Enzner, ``State-space architecture of the
  partitioned-block-based acoustic echo controller,'' in \emph{{IEEE}
  International Conference on Acoustics, Speech and Signal Processing, {ICASSP}
  2014, Florence, Italy, May 4-9, 2014}.\hskip 1em plus 0.5em minus 0.4em\relax
  {IEEE}, 2014, pp. 1295--1299. [Online]. Available:
  \url{https://doi.org/10.1109/ICASSP.2014.6853806}
\BIBentrySTDinterwordspacing

\bibitem{DBLP:conf/icml/KatharopoulosV020}
\BIBentryALTinterwordspacing
A.~Katharopoulos, A.~Vyas, N.~Pappas, and F.~Fleuret, ``Transformers are rnns:
  Fast autoregressive transformers with linear attention,'' in
  \emph{Proceedings of the 37th International Conference on Machine Learning,
  {ICML} 2020, 13-18 July 2020, Virtual Event}, ser. Proceedings of Machine
  Learning Research, vol. 119.\hskip 1em plus 0.5em minus 0.4em\relax {PMLR},
  2020, pp. 5156--5165. [Online]. Available:
  \url{http://proceedings.mlr.press/v119/katharopoulos20a.html}
\BIBentrySTDinterwordspacing

\bibitem{moore1983suggested}
B.~C. Moore and B.~R. Glasberg, ``Suggested formulae for calculating
  auditory-filter bandwidths and excitation patterns,'' \emph{The journal of
  the acoustical society of America}, vol.~74, no.~3, pp. 750--753, 1983.

\bibitem{stevens1937scale}
S.~S. Stevens, J.~Volkmann, and E.~B. Newman, ``A scale for the measurement of
  the psychological magnitude pitch,'' \emph{The journal of the acoustical
  society of america}, vol.~8, no.~3, pp. 185--190, 1937.

\bibitem{DBLP:conf/icassp/PanayotovCPK15}
\BIBentryALTinterwordspacing
V.~Panayotov, G.~Chen, D.~Povey, and S.~Khudanpur, ``Librispeech: An {ASR}
  corpus based on public domain audio books,'' in \emph{2015 {IEEE}
  International Conference on Acoustics, Speech and Signal Processing, {ICASSP}
  2015, South Brisbane, Queensland, Australia, April 19-24, 2015}.\hskip 1em
  plus 0.5em minus 0.4em\relax {IEEE}, 2015, pp. 5206--5210. [Online].
  Available: \url{https://doi.org/10.1109/ICASSP.2015.7178964}
\BIBentrySTDinterwordspacing

\bibitem{DBLP:conf/interspeech/ReddyDKNGCBGA021}
\BIBentryALTinterwordspacing
C.~K.~A. Reddy, H.~Dubey, K.~Koishida, A.~A. Nair, V.~Gopal, R.~Cutler,
  S.~Braun, H.~Gamper, R.~Aichner, and S.~Srinivasan, ``{INTERSPEECH} 2021 deep
  noise suppression challenge,'' in \emph{Interspeech 2021, 22nd Annual
  Conference of the International Speech Communication Association, Brno,
  Czechia, 30 August - 3 September 2021}, H.~Hermansky, H.~Cernock{\'{y}},
  L.~Burget, L.~Lamel, O.~Scharenborg, and P.~Motl{\'{\i}}cek, Eds.\hskip 1em
  plus 0.5em minus 0.4em\relax {ISCA}, 2021, pp. 2796--2800. [Online].
  Available: \url{https://doi.org/10.21437/Interspeech.2021-1609}
\BIBentrySTDinterwordspacing

\bibitem{DBLP:conf/icassp/KoPPSK17}
\BIBentryALTinterwordspacing
T.~Ko, V.~Peddinti, D.~Povey, M.~L. Seltzer, and S.~Khudanpur, ``A study on
  data augmentation of reverberant speech for robust speech recognition,'' in
  \emph{2017 {IEEE} International Conference on Acoustics, Speech and Signal
  Processing, {ICASSP} 2017, New Orleans, LA, USA, March 5-9, 2017}.\hskip 1em
  plus 0.5em minus 0.4em\relax {IEEE}, 2017, pp. 5220--5224. [Online].
  Available: \url{https://doi.org/10.1109/ICASSP.2017.7953152}
\BIBentrySTDinterwordspacing

\bibitem{Luo2019ConvTasNetSI}
Y.~Luo and N.~Mesgarani, ``Conv-tasnet: Surpassing ideal time-frequency
  magnitude masking for speech separation,'' \emph{{IEEE} {ACM} Trans. Audio
  Speech Lang. Process.}, vol.~27, no.~8, pp. 1256--1266, 2019.

\bibitem{DBLP:conf/icassp/RixBHH01}
\BIBentryALTinterwordspacing
A.~W. Rix, J.~G. Beerends, M.~P. Hollier, and A.~P. Hekstra, ``Perceptual
  evaluation of speech quality (pesq)-a new method for speech quality
  assessment of telephone networks and codecs,'' in \emph{{IEEE} International
  Conference on Acoustics, Speech, and Signal Processing, {ICASSP} 2001, 7-11
  May, 2001, Salt Palace Convention Center, Salt Lake City, Utah, USA,
  Proceedings}.\hskip 1em plus 0.5em minus 0.4em\relax {IEEE}, 2001, pp.
  749--752. [Online]. Available:
  \url{https://doi.org/10.1109/ICASSP.2001.941023}
\BIBentrySTDinterwordspacing

\bibitem{DBLP:conf/icassp/TaalHHJ10}
\BIBentryALTinterwordspacing
C.~H. Taal, R.~C. Hendriks, R.~Heusdens, and J.~Jensen, ``A short-time
  objective intelligibility measure for time-frequency weighted noisy speech,''
  in \emph{Proceedings of the {IEEE} International Conference on Acoustics,
  Speech, and Signal Processing, {ICASSP} 2010, 14-19 March 2010, Sheraton
  Dallas Hotel, Dallas, Texas, {USA}}.\hskip 1em plus 0.5em minus 0.4em\relax
  {IEEE}, 2010, pp. 4214--4217. [Online]. Available:
  \url{https://doi.org/10.1109/ICASSP.2010.5495701}
\BIBentrySTDinterwordspacing

\bibitem{DBLP:journals/sigpro/EnznerV06}
\BIBentryALTinterwordspacing
G.~Enzner and P.~Vary, ``Frequency-domain adaptive kalman filter for acoustic
  echo control in hands-free telephones,'' \emph{Signal Process.}, vol.~86,
  no.~6, pp. 1140--1156, 2006. [Online]. Available:
  \url{https://doi.org/10.1016/j.sigpro.2005.09.013}
\BIBentrySTDinterwordspacing

\end{thebibliography}

\end{document}